\documentclass[aps,prapplied,reprint]{revtex4-2}

\usepackage{graphicx}% Include figure files
\graphicspath{figures/}
\usepackage{dcolumn}% Align table columns on decimal point
\usepackage{amssymb}
\usepackage{amsmath}
\usepackage{fancyhdr}
\usepackage{bm}% bold math
\usepackage{times}
\usepackage{mathrsfs}
\usepackage[colorlinks,linkcolor=blue,anchorcolor=blue,citecolor=blue,urlcolor=blue]{hyperref}

\usepackage{changes}

\UseRawInputEncoding
\begin{document}
\title{Nonreciprocal interaction and entanglement between two superconducting qubits}
\author{Yu-Meng Ren}
\author{Xue-Feng Pan}
\author{Xiao-Yu Yao}
\author{Xiao-Wen Huo}
\author{Jun-Cong Zheng}
\author{Xin-Lei Hei}
\author{Yi-Fan Qiao}
\author{Peng-Bo Li}
\email{lipengbo@mail.xjtu.edu.cn}
\affiliation{Ministry of Education Key Laboratory for Nonequilibrium Synthesis and Modulation of Condensed Matter, Shaanxi Province Key Laboratory of Quantum Information and Quantum Optoelectronic Devices, School of Physics, Xi'an Jiaotong University, Xi'an 710049, China}

%\date{\today}

\begin{abstract}
Nonreciprocal interaction between two spatially separated subsystems plays a crucial role in signal processing and quantum networks. Here, we propose an efficient scheme to achieve nonreciprocal interaction and entanglement between two qubits by combining coherent and dissipative couplings in a superconducting platform, where two coherently coupled transmon qubits simultaneously interact with a transmission line waveguide. The coherent interaction between the transmon qubits can be achieved via capacitive coupling or via an intermediary cavity mode, while the dissipative interaction is induced by the transmission line via reservoir engineering. With high tunability of superconducting qubits, their positions along the transmission line can be adjusted to tune the dissipative coupling, enabling to tailor reciprocal and nonreciprocal interactions between the qubits. A fully nonreciprocal interaction can be achieved when the separation between the two qubits is $(4n+3)\lambda_{0} /4$, where $n$ is an integer and $\lambda_{0}$ is the photon wavelength. This nonreciprocal interaction enables the generation of nonreciprocal entanglement between the two transmon qubits. Furthermore, applying a drive field to one of the qubit can stabilize the system into a nonreciprocal steady-state entangled state. Remarkably, the nonreciprocal interaction in this work does not rely on the presence of nonlinearity or complex configurations, which has more potential applications in designing nonreciprocal quantum devices, processing quantum information, and building quantum networks.
\end{abstract}
\maketitle

\section{\label{sec:I}Introduction}
Reciprocity means that a physical system performs the same responses when sources and detectors are exchanged \cite{Jalas2013,Caloz2018,Nassar2020,Dong2021a}. By breaking this symmetry, nonreciprocity can be achieved in various ways, such as optomechanical interactions~\cite{Shen2018,Xu2019,Shen2023,Huang2024,Sun2024}, chiral quantum optics~\cite{Sayrin2015,Scheucher2016,Lodahl2017,Hurst2018,Zhang2018,Xia2018,Guddala2021,White2024} and hot atoms~\cite{Lin2019,Liang2020,Song2022,Zhang2023}. These approaches have also facilitated the development of nonreciprocal devices such as isolators~\cite{MercierdeLepinay2020,Ma2020,Kawaguchi2021,Sohn2021,Yu2023,White2023,Wu2024} and circulators~\cite{Rosenthal2017,Chapman2017,Wang2021,Herrmann2022,Navarathna2023,Fedorov2024}. The realization of nonreciprocity in the classical domain  can be categorized into three main groups: utilizing the Faraday effect in magnetic materials~\cite{Srinivasan2022,Yang2023,Yan2024}, applying temporal and spatial modulation~\cite{Sounas2017,Wang2018,Guo2019,Li2019,Peterson2019,Xu2020,Wang2020a,Chen2021,Wan2022,TessierBrothelande2023,Sisler2024}, and exploiting nonlinearity~\cite{RosarioHamann2018,Yang2019,Duggan2019,Guo2022}.

Recently, there has been growing interest in nonreciprocity within the quantum regime~\cite{Tang2022,Gueckelhorn2023,Shen2023a,Begg2024,Zhou2024,Reisenbauer2024,Li2024a}. By spinning the resonator to induce the Sagnac effect, various nonreciprocal quantum phenomena have been explored, including quantum entanglement~\cite{Jiao2020,Chen2023a}, photon and phonon blockade~\cite{Huang2018,Li2024,Yao2022,Zhang2023b}, phase transitions~\cite{Fruchart2021,Chiacchio2023,Zhu2024}, topological phonon transfer~\cite{lai2024nonreciprocal}, and multiquanta emission~\cite{Bin2024}. Another strategy utilizes an engineered reservoir to induce dissipative coupling~\cite{Metelmann2015,Fang2017,Lu2021b}, enabling nonreciprocity to arise from a balance between coherent and dissipative interactions~\cite{Wang2019,Huang2021,Pan2024}, thereby facilitating applications in quantum batteries~\cite{Ahmadi2024}. However, most existing researches have focused on nonreciprocal phenomena, predominantly within Bosonic systems.
The study of nonreciprocal interactions and entanglement between qubits within a simple framework remains largely unexplored, despite the fact that these interactions and entanglement are not only key components of quantum computing networks but also essential resources for one-way quantum communication~\cite{Wehner2018}.

As an emerging field focusing on the interaction between quantum emitters and waveguides with continuous electromagnetic field modes, waveguide quantum electrodynamics (wQED) has garnered significant interest in recent years~\cite{Xiang2013,Sheremet2023,Chang2018,Lemonde2018}. In the optical regime, these light-matter interaction paradigms have been experimentally realized in various physical systems, including atoms~\cite{GonzalezTudela2017,Corzo2019,Pennetta2022,Nie2023}, molecules~\cite{Tuerschmann2017}, quantum dots~\cite{Chu2023,Tiranov2023}, and spin defect centers~\cite{Burek2017,Bhaskar2017,Dong2021} coupled to optical waveguides. While at microwave frequencies, extensive investiagations have been conducted with superconducting qubits coupled to a transimssion line waveguide~\cite{Loo2013,Sundaresan2019,Aamir2022,Redchenko2023,Joshi2023}. As a promising candidate for quantum information processing, superconducting platforms offer significant potential for various quantum tasks~\cite{Nakamura1999,Kurpiers2018,Magnard2020,ruan2024dynamics,Zhang2023a,Wang2021a,Yin2023,Zhang2023c,Zhang2023d,Wang2024,Gu2024}, with superconducting qubits featuring advantages such as flexible scalability, in-situ tunability~\cite{Gu2017,Kannan2020}, and controllability~\cite{You2011}. Moreover, transmission lines can act as reservoirs, inducing controllable dissipative interactions between qubits through the continuum of electromagnetic field modes. Therefore, it is feasible to achieve both coherent and dissipative interactions simultaneously on superconducting quantum platforms.

In this work, we explore nonreciprocal interaction between two qubits by balancing coherent and dissipative couplings in a superconducting quantum circuit [Fig.~\ref{fig1:model}(a)], consisting of two transmon qubits and a transmission line. The coherent interaction can be achieved through a capacitance or via an intermediary cavity mode, while the dissipative interaction is introduced through reservoir engineering. With high tunability of superconducting platforms, the phase difference between the transmon qubits can be precisely tuned by altering their separation along the transmission line, which is essential for adjusting the dissipative coupling. This enables the competition between two types of couplings and tailors reciprocal and nonreciprocal interactions. Furthermore, our findings reveal that nonreciprocal interaction can induce novel quantum phenomena, such as nonreciprocal entanglement between two qubits. Under complete isolation, entanglement arises when one qubit is initially excited, but fails to form when the other qubit is excited, thus challenging the conventional perception that two identical qubits play the same role in forming a two-particle entangled state. Moreover, this nonreciprocal entanglement can be stabilized by applying a resonant drive to one of the qubit. This approach eliminates the need for magnetic fields, nonlinearity, or complex configurations, distinguishing it from previous methods and providing considerable potential as a promising candidate for generating nonreciprocal interaction and entanglement between qubits.

\section{\label{sec:II}The model}
\begin{figure*}[t]
	\centering
	\includegraphics[width=\textwidth]{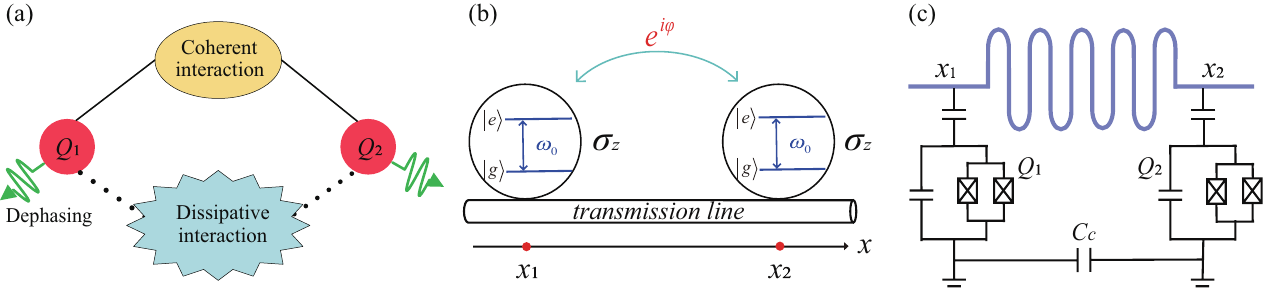}
	\caption{(a) Schematic of diverse interactions between two qubits. (b) A pair of superconducting qubits interacting with a one-dimensional transmission line. (c) Circuit implementation with two transmon qubits $Q_{1}$ (left, black) and $Q_{2}$ (right, black) capacitively coupled to a meandering transmission line (top, grey), resulting in dissipative coupling. Direct coherent coupling is achieved via a capacitance $C_{c}$.}
	\label{fig1:model}
\end{figure*}
As illustrated in Fig.~\ref{fig1:model}(b), two superconducting transmon qubits $Q_{1}$ and $Q_{2}$, with an energy level splitting $\omega_{0}$, are coupled to a transmission line at positions $x_1$ and $x_2$, respectively. Our goal is to realize a nonreciprocal interaction between two qubits by balancing coherent and dissipative interactions. For transmon qubits, the coherent interaction can be readily achieved via a capacitor $C_{c}$, as depicted in Fig.~\ref{fig1:model}(c). The Hamiltonian for coherent coupling takes the form ($\hbar = 1$)
\begin{equation}
	\hat{H}_{\text{coh}}=J \hat{\sigma}_{1}^{+} \hat{\sigma}_{2}^{-}+\text {H.c.}
\end{equation}
where $J$ denotes the direct coupling strength [see more details in Appendix~\ref{sec:appendixA}] and $\hat{\sigma}^{\pm}_{n}(n=1,2)$ are the raising and lowering operators, with subscripts 1 and 2 corresponding to $Q_{1}$ and $Q_{2}$, respectively. The total Hamiltonian of the system can be expressed as
\begin{equation}
	\hat{H}_{\text{total}}=\hat{H}_{\text{qb}}+\hat{H}_{\text{field}}+\hat{H}_{\mathrm{I}}.
\end{equation}
Here, the first term
\begin{equation}
\hat{H}_{\text{qb}}=\frac{\omega_{0}}{2} \sum_{n=1,2}\hat{\sigma}_{n}^{z}+\hat{H}_{\text{coh}}
\end{equation}
represents the free and coherent Hamiltonian of two qubits. The second term,
\begin{equation}\hat{H}_{\text {field}}=\sum_{q} \omega_{q} \hat{a}_{q}^{\dagger} \hat{a}_{q}
\end{equation}
corresponds to the Hamiltonian of the electromagnetic field modes, where $\omega_{q}$ is the frequency with wave vector $q$ and $\hat{a}_{q}^{\dagger}$~$(\hat{a}_{q})$ is the photon creation (annihilation) operator. The last term $\hat{H}_{\mathrm{I}}$ describes the interaction Hamiltonian between the qubits and the electromagnetic field in the waveguide, given by
\begin{equation}
\hat{H}_{\mathrm{I}}=\sum_{n=1,2}\left[e^{i \omega_{0} t} \hat{\sigma}_{n}^{-} \hat{E}\left(x_{n}, t\right)+\text { H.c.}\right],
\end{equation}
where
\begin{equation}
\hat{E}(x_n, t)=\sum_{q} g_{q}\left(\hat{a}_{q} e^{i q x_n-i \omega_{q} t}+\hat{a}_{q}^{\dagger} e^{-i q x_n+i \omega_{q} t}\right),
\end{equation}
and $g_{q}$ is the qubit-field coupling strength.

The dissipative interaction arises from continuous electromagnetic field modes in the transmission line. Applying the Born-Markov approximation and tracing out the photonic degrees of freedom, the dissipative part of the master equation, described with the reduced density operator $\hat{\rho}_S$, is given as
\begin{equation}
\mathcal{L}_e\left(\hat{\rho}_{\mathrm{S}}\right)=\sum_{n, m} J_{n, m}\left(\hat{\sigma}_{n}^{-} \hat{\rho}_{\mathrm{S}} \hat{\sigma}_{m}^{+}-\hat{\rho}_{\mathrm{S}} \hat{\sigma}_{m}^{+} \hat{\sigma}_{n}^{-}\right)+\text {H.c.}
\end{equation}
where the collective decay rates \(J_{n, m}\) are defined by~\cite{Gonzalez2013,Qiao2020}:
\begin{equation}
	J_{n, m}=\frac{\Gamma}{2} e^{i q\left(\omega_{0}\right)\left|x_{n}-x_{m}\right|}.
\end{equation}
Note that the collective decay rates of two qubits positioned at $x_1$ and $x_2$ depend on their separation $\Delta x =\left|x_{2}-x_{1}\right|$. The influence on phase in $J_{n, m}$ can be absorbed into the jump operator, enabling the control of dissipative interaction by varying $\varphi=2\pi\Delta x/\lambda_{0}$ with $\lambda_{0}=2\pi /q\left(\omega_{0}\right)$ the photon wavelength. Thus, the dissipative part of the Lindblad operator takes the form: $\mathcal{L}_e=\Gamma \mathcal{D}\left[\hat{\sigma}_{1}+e^{i \varphi} \hat{\sigma}_{2}\right] \hat{\rho}_{\mathrm{S}}$, where the superoperator is defined by $\mathcal{D}[\hat{o}] \hat{\rho}=\hat{o} \hat{\rho} \hat{o}^{\dagger}-\frac{1}{2} \hat{o}^{\dagger} \hat{o} \hat{\rho}-\frac{1}{2} \hat{\rho} \hat{o}^{\dagger} \hat{o}$. A detailed derivation is provided in Appendix~\ref{sec:appendixB}. Consequently, the evolution of two-qubits system is governed by the following master equation~\cite{Metelmann2015,Breuer2002,Reitz2022}:
\begin{equation}
	\begin{split}
		\frac{d}{d t} \hat{\rho}_{\mathrm{S}} = -i\left[\hat{H}_{\text{coh}}, \hat{\rho}_{\mathrm{S}}\right] + &\Gamma \mathcal{D}\left[\hat{\sigma}_{1}^{-}+e^{i \varphi} \hat{\sigma}_{2}^{-}\right] \hat{\rho}_{\mathrm{S}}\\
		+ \kappa \sum_{j \in 1,2} &\mathcal{D}\left[\hat{\sigma}_{j}^{z}\right] \hat{\rho}_{S}.
		\label{mastereq}
	\end{split}
\end{equation}
The first term in Eq.~(\ref{mastereq}) describes the direct coherent interaction between two qubits. In contrast, the second term denotes the dissipative coupling characterized by the decay rate $\Gamma$, where the phase difference $\varphi$ can be tuned to engineer collective effects between the qubits. The last term corresponds to the qubit dephasing with the rate $\kappa$.

\section{\label{sec:III}Nonreciprocal interaction}
\begin{figure}
	\centering
	\includegraphics{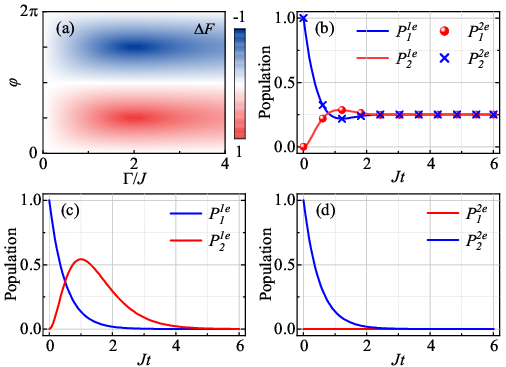}
	\caption{(a) Contour maps of the isolation ratio $\Delta F$ versus the ratio $\Gamma/J$ (dissipative to coherent coupling strength) and phase difference $\varphi$. Time evolution of the population for two qubits: (b) $\varphi = \pi$ (reciprocal case); (c) and (d) $\varphi=3 \pi/2$ (fully nonreciprocal case), with (c) only $Q_{1}$ initially excited, and (d) only $Q_{2}$ initially excited. The subscripts 1 and 2 in $P_i^{je}$ represent the populations of $Q_{1}$ and $Q_{2}$, respectively. While the superscripts $1e$ and $2e$ denote which qubit is initially excited. The parameters used in (b), (c), and (d) are $\Gamma=2J$, $\kappa=0$.}
	\label{fig:population-ge}
\end{figure}

To realize nonreciprocal interaction between the two qubits, a precise balance between the coherent and dissipative coupling is essential. The evolution of the spin operators, derived from Eq.~(\ref{mastereq}), is governed by the following equations:
\begin{equation}
	\begin{split}
		&\frac{d}{d t}\left\langle\hat{\sigma}_{1}^{-}\right\rangle = -\frac{\Gamma}{2}\left\langle\hat{\sigma}_{1}^{-}\right\rangle+\left(i J+\frac{\Gamma}{2} e^{i \varphi}\right)\left\langle\hat{\sigma}_{1}^{z}\hat{\sigma}_{2}^{-}\right\rangle,\\
		&\frac{d}{d t}\left\langle\hat{\sigma}_{2}^{-}\right\rangle = -\frac{\Gamma}{2}\left\langle\hat{\sigma}_{2}^{-}\right\rangle+\left(i J^{*}+\frac{\Gamma}{2} e^{-i \varphi}\right)\left\langle\hat{\sigma}_{2}^{z}\hat{\sigma}_{1}^{-}\right\rangle.
	\end{split}
	\label{eq:sigma-dynamics}
\end{equation}
In fact, the influence of engineered reservoir in the two spin operators is analogous to the effects of coherent interaction. The coefficients of the last term in Eq.~(\ref{eq:sigma-dynamics}) describe the effect from one qubit to the other, often referred to as the damping force ~\cite{Metelmann2015}. Here, we define the damping force as $F_{nm}$, indicating that the evolution of qubit $n$ depends on qubit $m$. To quantify the degree of nonreciprocal interaction, the isolation ratio $\Delta F$ is defined in terms of the damping force as
\begin{equation}
	\Delta F= \frac{\left|F_{12}\right|-\left|F_{21}\right|}{\left|F_{12}\right|+\left|F_{21}\right|},
	\label{DeltaF}
\end{equation}
where $\left|F_{12}\right|=\left|i J+\frac{\Gamma}{2} e^{i \varphi}\right|$ denotes how strongly $Q_{2}$ affects the dynamic of $Q_{1}$ while $\left|F_{21}\right|=\left|i J^{*}+\frac{\Gamma}{2} e^{-i \varphi}\right|$ denotes how strongly $Q_{2}$ is affected by $Q_{1}$. Obviously, the isolation ratio $\Delta F$ lies in the interval [-1,1]. Thus, Eq.~(\ref{DeltaF}) provides a metric for determining whether the interaction between the two qubits is nonreciprocal. When the isolation ratio $\Delta F=0$, we have $\left|F_{12}\right|=\left|F_{21}\right|$, which implies the interaction between the two qubits is reciprocal. In contrast, $\left|\Delta F\right|=1$ denotes the interaction is fully nonreciprocal. To be specific, $\Delta F =-1$ and $\Delta F =1$ correspond to unidirectional interactions from $Q_{1}$ to $Q_{2}$ and from $Q_{2}$ to $Q_{1}$, respectively. Other cases when $0~\textless~\left|\Delta F\right|\textless~1$ imply that the interaction between two qubits is nonreciprocal. Although both qubits can not be completely decoupled from the other qubit, the effects of the damping force on two qubits are not equal ($\left|F_{12}\right|\neq \left|F_{21}\right|$). The isolation ratio $\Delta F$ is then plotted as a function of the coupling strength ratio $\Gamma/J$ and the phase difference $\varphi$ in Fig.~\ref{fig:population-ge} (a). When $\Gamma=2J$, the isolation ratio reaches its minimum value of $\Delta F=-1$ at $\varphi=(4n+3) \pi/2$ and maximum value of $\Delta F=1$ at $\varphi=(4n+1) \pi/2$, corresponding to a full nonreciprocity. In addition, the reciprocal interaction occurs at $\varphi=n \pi$ ($n \in \mathbb{Z}$), i.e., $\Delta F=0$. A detailed discussion on the cases of full nonreciprocity and reciprocity follows.

Considering the case where only one qubit is initially in the excited state, we simulate the time evolution of the population expectation values $\left\langle\hat{\sigma}^{+} \hat{\sigma}^{-}\right\rangle$ for both qubits, as depicted in Figs.~\ref{fig:population-ge}(b), (c) and (d). We first examine the reciprocal case, $\Delta F=0$, where the separation between two qubits corresponds to $n \lambda_{0} / 2$ with $n \in \mathbb{Z}$, resulting in a phase difference of $\varphi = n\pi$. Under these conditions, the evolution of two qubits is identical, as illustrated in Fig.~\ref{fig:population-ge}(b). Next, we consider the case of fully nonreciprocal interaction from $Q_{1}$ to $Q_{2}$. By choosing
\begin{equation}
	J = i \frac{\Gamma}{2} e^{i \varphi}
	\label{isolatecondition},
\end{equation}
$\left\langle\hat{\sigma}_{1}^{-}\right\rangle$ is effectively decoupled from $\left\langle\hat{\sigma}_{2}^{-}\right\rangle$, consistent with the criterion of $\Delta F =-1$. This configuration ensures that the dynamics of  $\left\langle\hat{\sigma}_{1}^{-}\right\rangle$ is only confined to the first subsystem itself ($\left|F_{12}\right|=0$). In contrast, the time evolution of spin operators of $Q_{2}$ generally depends on both subsystems ($\left|F_{21}\right| \neq 0$). Thus, the interaction between two qubits is fully nonreciprocal, as $Q_{1}$ strongly affects $Q_{2}$ but remains unaffected by the other. Under complete isolation, the population dynamics of each qubit can be described as:
\begin{equation}
	\begin{split}
		&\frac{d}{dt}\left\langle\hat{\sigma}_1^{+}\hat{\sigma}_1^{-}\right\rangle=-\Gamma\left\langle\hat{\sigma}_1^{+}\hat{\sigma}_1^{-}\right\rangle\\
		&\frac{d}{dt}\left\langle\hat{\sigma}_2^{+}\hat{\sigma}_2^{-}\right\rangle=-\Gamma\left\langle\hat{\sigma}_2^{+}\hat{\sigma}_2^{-}\right\rangle-\Gamma e^{i\varphi}\langle\hat{\sigma}_1^{+}\hat{\sigma}_2^{-}\rangle-\Gamma e^{-i\varphi}\langle\hat{\sigma}_1^{-}\hat{\sigma}_2^{+}\rangle.
	\end{split}
\end{equation}
To achieve $\varphi=(4n+3) \pi/2$ with $n \in \mathbb{Z}$, the required separation is $(4 n+3) \lambda_{0} / 4$. Specifically, we set the coupling strength $\Gamma=2J$ and phase $\varphi=3 \pi/2$ to satisfy the condition of complete isolation in Eq.~(\ref{isolatecondition}). Consequently, $\Delta F=-1$ and the population dynamics of qubits exhibit distinct behaviors, as shown in Figs.~\ref{fig:population-ge}(c) and (d). When $Q_{1}$ is excited, $Q_{2}$ can receive excitation from $Q_{1}$, but $Q_{1}$ remains unaffected by exciting $Q_{2}$. Therefore, an unidirectional influence is established, where $Q_{2}$ is influenced by $Q_{1}$, but not vice versa. Alternatively, a separation of $(4 n+1) \lambda_{0} / 4$ enables unidirectional interaction from $Q_{2}$ to $Q_{1}$.

This configuration enables information to transfer unidirectionally between two qubits, while inhibiting in the reverse direction. The phase difference along the transmission line is determined by the separation between qubits. As the separation varies, different phases accumulate, causing the population dynamics of two qubits always evolve in different ways until the phase difference $\varphi = n\pi$. In other words, adjusting the phase difference facilitates the conversion between reciprocal and nonreciprocal interaction. It should be noted that, although qubit dephasing has been neglected in the above discussion, it does not affect the nonreciprocity between qubits. Because it has no influence on the nonreciprocal interaction, which is consistent with Eq.~(\ref{isolatecondition}).

\section{\label{sec:IV}Nonreciprocal entanglement}
\subsection{\label{sec:IVA}Nonreciprocal transient entanglement}
\begin{figure}
	\centering
	\includegraphics{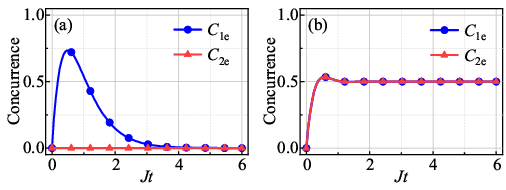}
	\caption{Time evolution of concurrence for the case of: (a) fully nonreciprocal transient entanglement with phase $\varphi=3\pi/2$, (b) reciprocal entanglement with phase $\varphi=0$. The subscript 1 or 2 in $C_{ie}$ represents the concurrence for initially excite $Q_{1}$ or $Q_{2}$. Other parameters used are $\Gamma=2J$, $\kappa=0$.}
	\label{fig:fig3-phase}
\end{figure}

So far, we have achieved nonreciprocal interaction between two qubits by balancing coherent and dissipative couplings. Building on this nonreciprocal interaction, we present a novel quantum phenomenon: the nonreciprocal transient entanglement.
To quantify the degree of entanglement between two qubits, we introduce concurrence $C$ defined by Wootters~\cite{Wootters1998}. For a separable state, the concurrence has the value $C = 0$, while for a maximally entangled state, $C=1$.
Similar to the discussion on population dynamics, we plot time evolution of concurrence with different separations between the qubits in Fig.~\ref{fig:fig3-phase}. The blue and red lines represent the concurrence when $Q_{1}$ or $Q_{2}$ is initially excited. When the separation is $(4 n+3) \lambda_{0} / 4$ [equivalent to the phase of $(4 n+3) \pi / 2$], the emergence of transient entanglement depends on which qubit is initially excited.

Specifically, the concurrence $C_{1e}$ evolves to a maximum value, while $C_{2e}$ remains at 0, indicating no entanglement occurs when $Q_{2}$ is excited, as shown in Fig.~\ref{fig:fig3-phase}(a).
Although other parameters remain the same, the dynamics of concurrence exhibit different behaviors depending on which qubit is initially excited. This difference indicates that the entanglement is influenced by the nonreciprocal interaction. Physically, this phenomenon can be explained as follows: due to both coherent and dissipative interactions, $Q_{2}$ receives excitation transferred from $Q_{1}$, resulting in a transient entanglement between them. In contract, when the other qubit $Q_{2}$ is excited and the parameters satisfied the criterion for $\Delta F =-1$, the damping force from $Q_{2}$ to $Q_{1}$ is $\left|F_{12}\right|=0$; thus the evolution of $Q_{1}$ remains completely independent of $Q_{2}$. However, the entanglement resulting from two different initial states can exhibit identical behavior by adjusting the separation to $n \lambda_{0} / 2$, as shown in Fig.~\ref{fig:fig3-phase}(b).

\subsection{\label{sec:IVB}Nonreciprocal stabilized entanglement}
Furthermore, we consider the case where a drive is applied to a qubit to provide a continuous excitation. The driving term is given by $\hat{H}_\text{drive} = \Omega_{d}(\hat{\sigma}_{i}^{+} e^{-i \omega_{d} t} + \hat{\sigma}_{i}^{-} e^{i \omega_{d} t})$ with the drive strength $\Omega_{d}$ and frequency $\omega_d$. By rotating the reference frame and setting $\omega_{d} = \omega_{0}$ to achieve resonance, we obtain the time-independent driving term in the interaction picture: $\hat{H}_\text{drive}^{I} = \Omega_{d}(\hat{\sigma}^{+}+\hat{\sigma}^{-})$. Therefore, the master equation [Eq.~\ref{mastereq}] becomes:
\begin{equation}
	\begin{split}
		\frac{d}{d t} \hat{\rho}_{\mathrm{S}} = -i \left [\hat{H}_{\text{coh}}+\hat{H}_{\text{drive}}^{I}, \hat{\rho}_{\mathrm{S}}\right] + \Gamma &\mathcal{D}\left[\hat{\sigma}_{1}^{-}+e^{i \varphi} \hat{\sigma}_{2}^{-}\right] \hat{\rho}_{\mathrm{S}}\\
		+ \kappa \sum_{j \in 1,2} \mathcal{D}\left[\hat{\sigma}_{j}^{z}\right] &\hat{\rho}_{S}
	\end{split}
	\label{mastereq+drive}
\end{equation}

\begin{figure}
	\centering
	\includegraphics{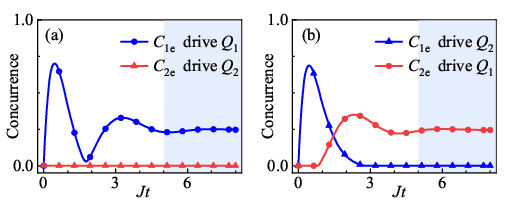}
	\caption{Concurrence versus time when a driving term is applied to (a) the initially excited qubit, (b) the other qubit which is initially in the ground state. The $C_{1e}$ denotes the concurrence for whose initial state is $\left | \psi_{0}  \right \rangle =\left | e \right \rangle_{1} \left |g \right \rangle_{2}$ while $C_{2e}$ for $\left | \psi_{0}  \right \rangle =\left | g \right \rangle_{1} \left |e \right \rangle_{2}$. The parameters used are $\Omega_{d}\sim 8J/11$, in the complete isolation and no dephasing $\kappa=0$.}
	\label{fig:drive}
\end{figure}

When the drive is applied, the transient entanglement eventually evolves to a stable value, as depicted in the blue shaded regions of Fig.~\ref{fig:drive}. Remarkably, this stabilized entanglement exhibits nonreciprocal characteristics. The blue line in Fig.~\ref{fig:drive}(a) shows that when $Q_{1}$ is initially excited and the drive is applied, the concurrence $C_{1e}$ rises from 0 to a maximum value and eventually stabilizes. In contrast, no entanglement occurs when the drive is applied to the other initially excited qubit, $Q_{2}$, as shown by the red line in Fig.~\ref{fig:drive}(a). Another scenario we consider is when the drive is applied to a qubit that is initially in the ground state, as illustrated in Fig.~\ref{fig:drive}(b). The red line in Fig.~\ref{fig:drive}(b) depicts the case where $Q_{2}$ is excited initially while $Q_{1}$ is driven. Although no entanglement is present initially, the concurrence gradually emerges over time and eventually stabilizes. Conversely, when the drive is applied to $Q_{2}$, the concurrence decays to 0 regardless of how $Q_{2}$ evolves, as the population of $Q_{1}$ decays over time. This behavior arises because, when the isolation ratio $\Delta F=-1$, the evolution of $Q_{1}$ remains entirely independent of $Q_{2}$, whereas $Q_{2}$ depends on both qubits. Comparing Fig.~\ref{fig:drive}(a) and (b), we conclude that once the direction of fully nonreciprocal interaction is established, nonreciprocal stabilized entanglement can be achieved by driving the qubit that evolves independently, regardless of the initial state.

\section{\label{sec:V}Collective state basis}
\begin{figure}
	\centering
	\includegraphics{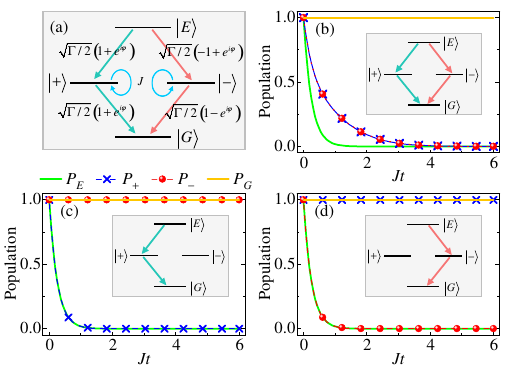}
	\caption{(a) Effective processes of the system. The coherent Hamiltonian cause the interactions to itself with an effective strength $J$. The qubits decay through the transmission-line via two paths: one from $\left | E \right \rangle$ to $\left | + \right \rangle$ and from $\left | + \right \rangle$ to $\left | G \right \rangle$ with an effective decay rate of $\sqrt{\Gamma/2}(1+e^{i\varphi})$; the other from $\left | E \right \rangle$ to $\left | - \right \rangle$ and from $\left | -  \right \rangle$ to $\left | G \right \rangle$ with effective decay rates of  $\sqrt{\Gamma/2}(-1+e^{i\varphi})$ and $\sqrt{\Gamma/2}(1-e^{i\varphi})$, respectively. Time evolution of the respective population $P_{E}$, $P_{+}$, $P_{-}$, and $P_{G}$ from different initial states $\left | E \right \rangle$, $\left | + \right \rangle$, $\left | - \right \rangle$, and $\left | G \right \rangle$ under different phase difference (b) $\varphi=3 \pi/2$, (c) $\varphi=0 $, (d) $\varphi=\pi$. The parameters used are $\Gamma=2J$ and $\kappa=0$.}
	\label{fig:popu-E+-G}
\end{figure}

To have a better understanding of the nonreciprocal dynamics, we introduce the collective state basis for the two transmon qubits: $\{\left | E \right \rangle = \left | e \right \rangle_{1} \left |e \right \rangle_{2}, \left | \pm \right \rangle = (1/\sqrt{2})(\left | e \right \rangle_{1} \left |g \right \rangle_{2}\pm\left | g \right \rangle_{1} \left |e \right \rangle_{2}), \left | G \right \rangle = \left| g \right \rangle_{1} \left|g \right \rangle_{2}\}$. In this basis, the coherent Hamiltonian $\hat{H}_{\text{coh}}$ and the Lindblad operator $\mathcal{L}_e=\Gamma \mathcal{D}\left[\hat{\sigma}_{1}^{-}+e^{i \varphi} \hat{\sigma}_{2}^{-}\right]$ are expressed as:
\begin{equation}
	\mathscr{H}_{\text{coh}}=J \left(\left| + \right \rangle \left\langle + \right| - \left| - \right \rangle \left\langle - \right| \right),
\end{equation}
\begin{equation}
	\begin{split}
		\mathscr{L}_e = \sqrt{\Gamma/2} \left[
		(1+e^{i\varphi}) \left| + \right\rangle \left\langle E \right| + (1+e^{i\varphi}) \left| G \right\rangle \left\langle + \right| \right. \\
		\left. + (-1+e^{i\varphi}) \left| - \right\rangle \left\langle E \right| + (1-e^{i\varphi}) \left| G \right\rangle \left\langle - \right| \right].
	\end{split}
\end{equation}
The effective processes of how coherent and dissipative interactions influence the two qubits are illustrated in Fig.~\ref{fig:popu-E+-G}(a). The coherent interaction does not cause the two qubits to decay. While the dissipative interaction induces decay through two paths with different rates. Specifically, when the phase is $\varphi=(2n+1) \pi/2$ (the complete isolation), two decay paths are identical, and their decay rates have the same magnitude. If the phase is set to $\varphi=n \pi$, one of the collective states $\left | + \right \rangle$ or $\left | - \right \rangle$ can be completely decoupled from the dynamics of the remaining three states depending on $n$ is even or odd, as depicted in Fig.~\ref{fig:popu-E+-G}(c) and (d).

\begin{figure}
	\centering
	\includegraphics{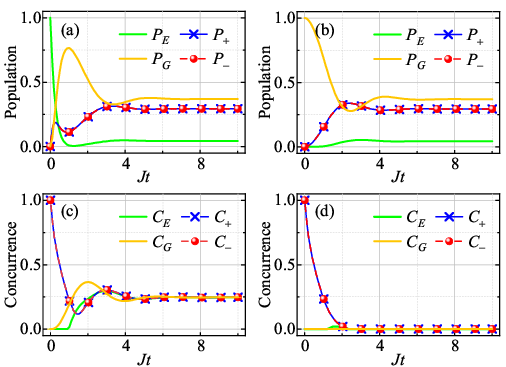}
	\caption{Time evolution of the respective population $P_{E}$, $P_{+}$, $P_{-}$, $P_{G}$ with the drive applied to $Q_{1}$, starting from two different initial states (a) $\left | E \right \rangle$, (b) $\left | G \right \rangle$. Time evolution of the respective concurrence $C_{E}$, $C_{+}$, $C_{-}$, and $C_{G}$ from different initial states $\left | E \right \rangle$, $\left | + \right \rangle$, $\left | - \right \rangle$, and $\left | G \right \rangle$ with the same drive applied to different qubits: (c) driving $Q_{1}$, (d) driving $Q_{2}$. The parameters used are $\Omega_{d}\sim 8J/11$, in the complete isolation ($\varphi=3 \pi/2$ and $\Gamma=2J$), and $\kappa=0$.}
	\label{fig:Con-E+-G}
\end{figure}

So far, the influences of coherent and dissipative Hamiltonian have been discussed above. Now we consider the case with an additional driving term.
Fig.~\ref{fig:Con-E+-G}(a) and (b) depict the dynamics of the respective population of collective states $P_{E}$, $P_{+}$, $P_{-}$, and $P_{G}$ when applying the driving $\Omega_{d}$ on $Q_{1}$, starting from two different initial states $\left | E \right \rangle$ and $\left | G \right \rangle$. Notably, the populations of collective states eventually reach specific values, independent of the choice of initial states. In other words, the system evolves to the same stationary state irrespective of the initial states. Furthermore, we simulate the concurrence between two qubits from different initial states $\left | E \right \rangle$, $\left | + \right \rangle$, $\left | - \right \rangle$, and $\left | G \right \rangle$. When $Q_{1}$ is driven with a strength of  $\Omega_{d}$ under complete isolation, the corresponding concurrence converges to the same value, independent of the initial states, as shown in Fig.~\ref{fig:Con-E+-G}(c). However, when the drive is applied on $Q_{2}$ using the same parameters, the concurrence always decays to 0 eventually, indicating that fully nonreciprocal entanglement is achieved by driving a different qubit.

\section{\label{sec:VI}Experimental feasibility}
We consider the  experimental feasibility of this scheme using superconducting quantum circuits. Within the framework of wQED, it is readily accessible to achieve qubit-waveguide coupling efficiencies exceeding 99$\%$~\cite{Mirhosseini2019,Kannan2020a,Kannan2023}. The specific implementation consists of two transmon qubits, capacitively coupled to a microwave transmission line with a characteristic impedance of $Z_{0}\simeq 50~\Omega$~\cite{Redchenko2023}. The separation of two qubits along the waveguide can also be adjusted by tuning the qubit frequency. Specifically, $Q_{1}$ and $Q_{2}$ are tuned to a frequency of 4.8~$\mathrm{GHz}$ or 6.4 $\mathrm{GHz}$, corresponding to a separation of $3\lambda_{0} / 4$ or $\lambda_{0}$ with $\lambda_{0}$ the photon wavelength, respectively~\cite{Loo2013}. The waveguide's lateral dimension is on the order of 10 $\mu$m while its wavelength is on the order of 10 mm~\cite{Roy2017}. These components are typically patterned on sapphire substrates~\cite{Satzinger2018}. Recent experiment has achieved the value of relaxation time $T_{1}$ up to 500 $\mu$s for tantalum-based transmon qubits using a dry etching process~\cite{Wang2022}. Additionally, millisecond coherence times are attainable via optimization of materials and circuit geometry modifications~\cite{Ganjam2024}. It has been demonstrated in experiment that the relaxing rate between transmon qubit and transmission line can be set at $\Gamma=100 ~\mathrm{MHz}$~\cite{Mirhosseini2019}. Due to the tunability of capacitive coupling between the two qubits, the coherent coupling strength can be adjusted to $J=\Gamma/2=50~ \mathrm{MHz}$. From Fig.~\ref{fig:drive}, the time for our system to reach the stabilized entanglement is about $Jt \lesssim 5$, i.e., $T \approx 130~\mu s$, which is much shorter than the coherence time.
The decoherence rate of a qubit, $\Gamma_\mathrm{tol}=\Gamma/2+\Gamma_\mathrm{nr}$, includes contributions from radiative coupling $\Gamma$ and non-radiative decoherence rate $\Gamma_\mathrm{nr} = \Gamma'_\mathrm{nr}/2+\kappa_{\phi}$ with non-radiative energy loss $\Gamma'_\mathrm{nr}$ and pure dephasing $\kappa_{\phi}$~\cite{Zanner2022}.
By measuring the waveguide transmission or reflection in scattering experiments, both the coupling rate to the waveguide $\Gamma$ and non-radiative decoherence rate $\Gamma_\mathrm{nr}$ can be determined~\cite{Lu2021a}. In our setup, parasitic decoherence rate due to damping and dephasing to other channels apart from waveguide is on the order of 100 $\mathrm{kHz}$~\cite{Mirhosseini2019}, which is much smaller than the coupling rate and can be estimated. Here, we select transmon as qubits in our model for its prevalence and the flexibility of cQED architectures. Additionally, transmon qubits offer a reduced complexity and fewer new decoherence sources compared to other different elements~\cite{Huang2020}, making them particularly suitable for our setup.

\section{\label{sec:VII}Conclusion}
We propose how to achieve the nonreciprocal interaction between two quantum two-level systems on a superconducting platform and explore how this nonreciprocity results in the nonreciprocal entanglement. In our scheme, the nonreciprocal interaction is achieved through a balance of coherent coupling via a capacitor, and dissipative coupling via a transmission line which provides the engineered reservoir. The high tunability of qubit parameters, such as their positions along the transmission line and their frequencies, enables the design of tailored reciprocal and nonreciprocal interactions between qubits. A fully nonreciprocal interaction is achieved with a separation of $(4n+3) \lambda_{0} / 4$, where $\lambda_{0}$ is the photon wavelength. To quantify the degree of isolation, we introduce a criterion to identify the nonreciprocal interaction. Induced by nonreciprocal interaction, a fully nonreicprocal transient entanglement emerges unidirectionally, existing when $Q_{1}$ is excited while vanishing when the other qubit is excited. Furthermore, by applying a drive to the qubit, nonreciprocal stabilized entanglement can also be achieved through the joint action of three types of interactions. We also explain the aforementioned nonreciprocity with the collective state basis, providing a means to deepen our understanding of these quantum nonreicprocal interactions. Our scheme offers a potential pathway for investigating directional quantum information transmission and developing designs for one-way quantum devices~\cite{Guimond2020,Shen2023,Zhang2021a,Wang2022b}.

\begin{acknowledgments}
P.B.L. is supported by the National Natural Science
Foundation of China under Grants No. 12375018 and No. W2411002.
\end{acknowledgments}

\appendix
\section{\label{sec:appendixA}The coherent coupling}
Now we consider the concrete form of two transmon qubits coherent couplings. For each transmon qubit, it can be described as:
\begin{equation}
	\hat{H}_{\text{transmon}}=4E_C(\hat{n}-n_g)^2-E_J\cos(\hat{\varphi})
	\label{HT0}
\end{equation}
Here, \(E_{C}=e^{2}/2C_{\Sigma}\) denotes the
charging energy, with \(C_\Sigma=C_J+C_S\) the total capacitance including
$C_J$ the capacitance of junction and $C_S$ the shunt capacitance. The operator $\hat{n}=\hat{Q}/2e$ denotes the charge number, while operator $\hat{\varphi}=(2\pi/\Phi_0)\hat{\Phi}$ denotes the phase and $n_g=Q_g/2e$ for a possible offset charge. In the transmon regime, the frequency of the first energy level is insensitive to the variations in the offset charge $n_g=Q_g/2e$ thus it is ignored in the subsequent discussion. It is useful to introduce creation
and annihilation operators ($\hat{b}^\dagger$ and $\hat{b}$) for quantization in
analogy to the quantization of harmonic oscillators; thus the phase and charge operators are given by $	\hat{\varphi}=\left(2E_{C}/{E_{J}}\right)^{1/4}(\hat{b}^{\dagger}+\hat{b})$, and $\hat{n}={i}\left(E_{J}/{2E_{C}}\right)^{1/4}(\hat{b}^{\dagger}-\hat{b})/{2}$. Therefore, in a frame rotating at $\omega_{t}$, Eq.~(\ref{HT0}) has the form $\hat{H}_{\text{transmon}}
\approx\hbar\omega_{t}\hat{b}^{\dagger}\hat{b}-\frac{E_{C}}{2}\hat{b}^{\dagger}\hat{b}^{\dagger}\hat{b} \hat{b}$,
with $\hbar\omega_{t}=\sqrt{8E_{C}E_{J}}-E_{C}$.
For two transmon qubits via a capacitive coupling, with the Hamiltonian of two qubits described as $\hat{H}_1$ and $\hat{H}_2$ respectively, the whole Hamiltonian can be described as
\begin{equation}
	\hat{H}=\hat{H}_1+\hat{H}_2+\hat{H}_{\text{coupling}}.
\end{equation}
To be specific, the two transmon qubits have the following forms of Hamiltonian
\begin{equation}
	\begin{split}
		&\hat{H}_1=\hbar\omega_{t1}\hat{b}_1^{\dagger}\hat{b}_1-\frac{E_{C_1}}{2}\hat{b}_1^{\dagger}\hat{b}_1^{\dagger}\hat{b}_1 \hat{b}_1\\
		&\hat{H}_2=\hbar\omega_{t2}\hat{b}_2^{\dagger}\hat{b}_2-\frac{E_{C_2}}{2}\hat{b}_2^{\dagger}\hat{b}_2^{\dagger}\hat{b}_2 \hat{b}_2
	\end{split}
\end{equation}
where $\omega_{ti}$ denotes the frequency of $i$th transmon qubit, $\hat{b}_i$ ($\hat{b}^{\dagger}_i$) denotes the annihilation (creation) operators.
And the coupling term, arising from the capacitor $C_{c}$, can be written as~\cite{Blais2021}
\begin{equation}
	\hat{H}_{\text{coupling}}=\frac{2E_{C1}E_{C2}}{E_{C_{c}}}\left(\frac{E_{J1}}{2E_{C1}}\times\frac{E_{J2}}{2E_{C2}}\right)^{1/4}(\hat{\sigma}_{1}^{+} \hat{\sigma}_{2}^{-}+\hat{\sigma}_{1}^{-} \hat{\sigma}_{2}^{+})
\end{equation}
where $E_{C1} (E_{C2})$ and $E_{J1} (E_{J2})$ are the charging and Josephson energies of two transmon qubits, and $E_{C_{c}}=e^2/2C_{c}$ is the charging energy of the capacitance $C_{c}$. Supposing that the two transmon qubits are tuned in resonance and in a frame rotating at the frequency of qubits $\omega_0=\omega_{t1}=\omega_{t2}$, the coupling term takes the form
\begin{equation}
	\hat{H}_{\text{coh}}=J(\hat{\sigma}_{1}^{+} \hat{\sigma}_{2}^{-}+\hat{\sigma}_{1}^{-} \hat{\sigma}_{2}^{+})
\end{equation}

\section{\label{sec:appendixB}The dissipative coupling}
In this section, we provide a detailed derivation of the dissipative part of the Lindblad operator. Since collective dissipation cannot be directly represented by a Hamiltonian, we apply the Lindblad master equation formalism to model the dissipative interaction induced by the transmission line. One can assume that the coupling between the subsystems and the reservoir is weak (Born approximation), and that the correlation times of the reservoir are much shorter than the characteristic timescales of the system, thus making the process Markovian~\cite{scully1997quantum}. After tracing out the photonic degrees of freedom, the dynamics of the reduced density operator $\hat{\rho} _{S}$ for the qubits are generally formulated as~\cite{Breuer2002,Reitz2022}
\begin{eqnarray}
\frac{d}{d t} \hat{\rho}_{\mathrm{S}}(t)&=&-\int_{0}^{\infty} d s \operatorname{Tr}_{\mathrm{R}}\left\{\left[\hat{H}_{\text{int}}(t),\right.
\left.\left[\hat{H}_{\text{int}}(t-s),\hat{\rho}_{\mathrm{S}}(t) \otimes \hat{\rho}_{\mathrm{R}}\right]\right]\right\}
\label{eq:dynamic ps}
\end{eqnarray}
where $\hat{\rho} _{S}=\operatorname{Tr}_{\mathrm{R}}( \hat{\rho})$ with subscript $S$ for qubits and $R$ for the electromagnetic field.
The dipolar coupling between qubits and the transmission line $\hat{H}_{\mathrm{I}}=\sum_{n=1,2}\left[\hat{\sigma}_{n}\hat{E}\left(x_{n}\right)+\text{~H.c.}\right]$ with $\hat{E}(x)=\sum_{q}g_{q}\left(\hat{a}_{q}e^{iqx}+\hat{a}_{q}^{\dagger}e^{-iqx}\right)$, can be transformed and subsequently expressed in the interaction picture as:
\begin{equation}
	\hat{H}_{\mathrm{I}}^{\mathrm{I}}=\sum_{n=1,2}\left[e^{i \omega_{0} t} \hat{\sigma}_{n} \hat{E}\left(x_{n}, t\right)+\text {H.c.}\right]
	\label{HI}
\end{equation}
with $\hat{E}(x, t)=\sum_{q} g_{q}\left(\hat{a}_{q} e^{i q x-i \omega_{q} t}+\hat{a}_{q}^{\dagger} e^{-i q x+i \omega_{q} t}\right)$ and $g_{q}$ the qubit-field coupling strength.

Substituting the specific form of interaction Hamiltonian $\hat{H}_{\mathrm{I}}^{\mathrm{I}}$ into Eq.~(\ref{eq:dynamic ps}), the Lindblad operator for the dissipative part is given by:
\begin{equation}
	\mathcal{L}_e\left(\hat{\rho}_{\mathrm{S}}\right)=\sum_{n, m} J_{n, m}\left(\hat{\sigma}_{n}^{-} \hat{\rho}_{\mathrm{S}} \hat{\sigma}_{m}^{+}-\hat{\rho}_{\mathrm{S}} \hat{\sigma}_{m}^{+} \hat{\sigma}_{n}^{-}\right)+\text {H.c.}
\end{equation}
Here, the collective decay rates $J_{n, m}$ are defined as
\begin{equation}
	J_{n, m}=\sum_{q}\left|g_{q}\right|^{2} \int_{0}^{\infty} d s\left(e^{-i \omega_{0} s}+e^{i \omega_{0} s}\right) e^{-i \omega_{q} s} \\\cdot e^{i q\left|x_{n}-x_{m}\right|},
\end{equation}
assuming the field is in vacuum, thus satisfying $\left \langle \hat{a}_{q}^{\dagger} \hat{a}_{q} \right \rangle = 0$. To derive an explicit expression for $J_{n, m}$, we perform additional calculations and make specific assumptions. In the continuum limit, we approximate the series sum as an integral and apply the dispersion relation for 1D waveguide, $\omega_{q} = cq$. Consequently, the collective decay rates are simplified to~\cite{Gonzalez2013,Qiao2020}:
\begin{equation}
	J_{n, m}=\frac{\Gamma}{2} e^{i q\left(\omega_{0}\right)\left|x_{n}-x_{m}\right|}.
	\label{Jnm}
\end{equation}
Here we define $\Gamma=\gamma\left(\omega_{0}\right)$, with the function $\gamma(\omega)=g_{q(\omega)}^{2} D(\omega) / \pi$ and the electromagnetic field state density $D(\omega)=(2 \pi / L)|d q(\omega) / d \omega|$, with $L$ the quantization length.

The collective decay rates $J_{n, m}$ depend on the separation $\Delta x =\left|x_{n}-x_{m}\right|$ between the two qubits, enabling the dissipative interaction to be easily tuned by adjusting their positions. To simplify the exponential part, we introduce $i \varphi$ to replace the $i q\left(\omega_{0}\right)\left|x_{n}-x_{m}\right|$ in Eq.~(\ref{Jnm}), where $\varphi=2\pi\left|x_{n}-x_{m}\right|/\lambda_{0}$ is defined by the photon wavelength $\lambda_{0}=2\pi/q\left(\omega_{0}\right)$. Furthermore, the specific formulations of collective decay rates, such as $J_{11}$ and $J_{12}$ are expressed with $\varphi$.  Moreover, the phase effect of the collective decay rates $J_{n, m}$ can be absorbed into the jump operator, resulting in the following specific form for the dissipative interaction
\begin{equation}
	\mathcal{L}_e=\Gamma \mathcal{D}\left[\hat{\sigma}_{1}+e^{i \varphi} \hat{\sigma}_{2}\right] \hat{\rho}_{\mathrm{S}}.
\end{equation}

%\bibliography{chugao-ref}


\end{document}